\documentclass[graybox,natbib,nosecnum]{svmult}
\bibpunct{(}{)}{;}{a}{}{,} 

\pdfoutput=1   

\usepackage{mathptmx}       
\usepackage{helvet}         
\usepackage{courier}        
\usepackage{type1cm}        

\usepackage{makeidx}         
\usepackage{graphicx}        
\usepackage{multicol}        
\usepackage[bottom]{footmisc}
\usepackage[normalem]{ulem}	
\usepackage{hyperref}  

\usepackage{soul}   

\newcommand*\aap{A\&A}

\newcommand*\aj{AJ}

\newcommand*\apj{ApJ}
\newcommand*\apjl{ApJ}

\newcommand*\araa{ARA\&A}

\newcommand*\icarus{Icarus}

\newcommand*\jgr{J Geophys Res}

\newcommand*\mnras{MNRAS}

\newcommand*\nat{Nature}

\newcommand{\ale}{\ \raisebox{-.3ex}{$\stackrel{<}{\scriptstyle \sim}$}\ }
\newcommand{\age}{\ \raisebox{-.3ex}{$\stackrel{>}{\scriptstyle \sim}$}\ }

\makeindex             

\begin{document}

\title*{A Brief Overview of Planet Formation}
\author{Philip J. Armitage}
\institute{JILA, University of Colorado \& NIST,  UCB 440, Boulder, CO 80309-0440, USA,
\email{pja@jilau1.colorado.edu}}
%
%
\maketitle

\abstract{The initial conditions, physics, and outcome of planet formation are now constrained by 
detailed observations of protoplanetary disks, laboratory experiments, and the discovery of thousands 
of extrasolar planetary systems. These developments have broadened the range of processes 
that are considered important in planet formation, to include disk turbulence, radial drift, planet migration, 
and pervasive post-formation dynamical evolution. The N-body collisional growth of planetesimals and 
protoplanets, and the physics of planetary envelopes---key ingredients of the classical model---remain 
central. I provide an overview of the current status of planet formation theory, and 
discuss how it connects to observations.}

\section{Introduction}
Solar System and astronomical evidence of the origin of planets is most naturally interpreted in terms of a 
bottom-up theory \citep{safronov72}, in which planetary systems form within 
largely gaseous protoplanetary disks from initially microscopic solid material. Different 
physical processes dominate as growth proceeds. The earliest phases (corresponding to 
particle sizes of $s \sim \mu$m-m) involve primarily aerodynamic and material physics. Gravitational 
forces become increasingly important later on, first between growing planetesimals  
($s \age {\rm km}$) and later between protoplanets and gas in the disk 
(for masses $M \age 0.1 \ M_\oplus$, where $M_\oplus$ is the mass of the Earth). Giant planet growth from $\sim 3-20+ \ M_\oplus$ 
{\em cores} is limited initially by the ability of their gaseous envelopes to cool, and subsequently 
by how fast the surrounding disk can supply mass. Finally, the planetary systems that we 
observe---often after an interval of several Gyr---can be profoundly 
modified from their initial state by dynamical instabilities, secular evolution, and tides.

This review is an introduction to the processes that matter 
during planet formation, how those processes may combine to yield planetary systems, and 
where the theory can be tested against observations. To the extent that there is a 
theme, it is {\em mobility}---of gas in the disk, of dust and pebbles under 
aerodynamic forces, and of planets due to gravitational torques against the gas and interactions 
with other bodies. Mobility, particularly in the guise of the radial drift of 
particles or the migration of low-mass planets through gaseous disks, was once seen as a ``problem" to be 
ideally solved or otherwise ignored. The current view is more positive. Mobility 
is due to clearly defined physical processes, opens up new routes for rapid growth, and is key 
to the architecture of many observed extrasolar planetary systems.

\section{Protoplanetary disks}
The kinematics of protoplanetary disks, their thermal and chemical structures, and their 
evolutionary histories are all key to planet formation. Most attention focuses on Class II 
Young Stellar Objects (YSOs) \citep{lada87}, when the star has attained close to its final mass 
and the disk is low mass ($M_{\rm disk} \ll M_*$) and relatively long-lived (several Myr). (It remains 
possible, however, that significant particle growth occurs during prior embedded phases.) Observational 
inferences of the stellar mass accretion rate $\dot{M}$ in Class II sources are moderately robust, and typically 
yield $\dot{M} \sim 10^{-8.5} \ M_\odot \ {\rm yr}^{-1}$ for $M_* \approx M_\odot$, with a super-linear scaling 
with stellar mass \citep{alcala17}. Disk mass estimates are problematic, because H$_2$ is not detected 
directly. For a very small number of disks \citep[including TW~Hya;][]{bergin13} HD emission in the far-infrared 
has been observed, and mass estimates based on this tracer \citep{trapman17} provide calibration for more 
accessible estimators.  Estimates based on scaling the mm continuum emission from dust yield a median ratio 
$M_{\rm disk} / M_* \simeq 10^{-2.5}$ for disks in Taurus \citep{andrews13}. Modeling of CO isotopologue 
line emission gives on average lower values \citep{williams14}.

\begin{figure}
\includegraphics[width=\textwidth]{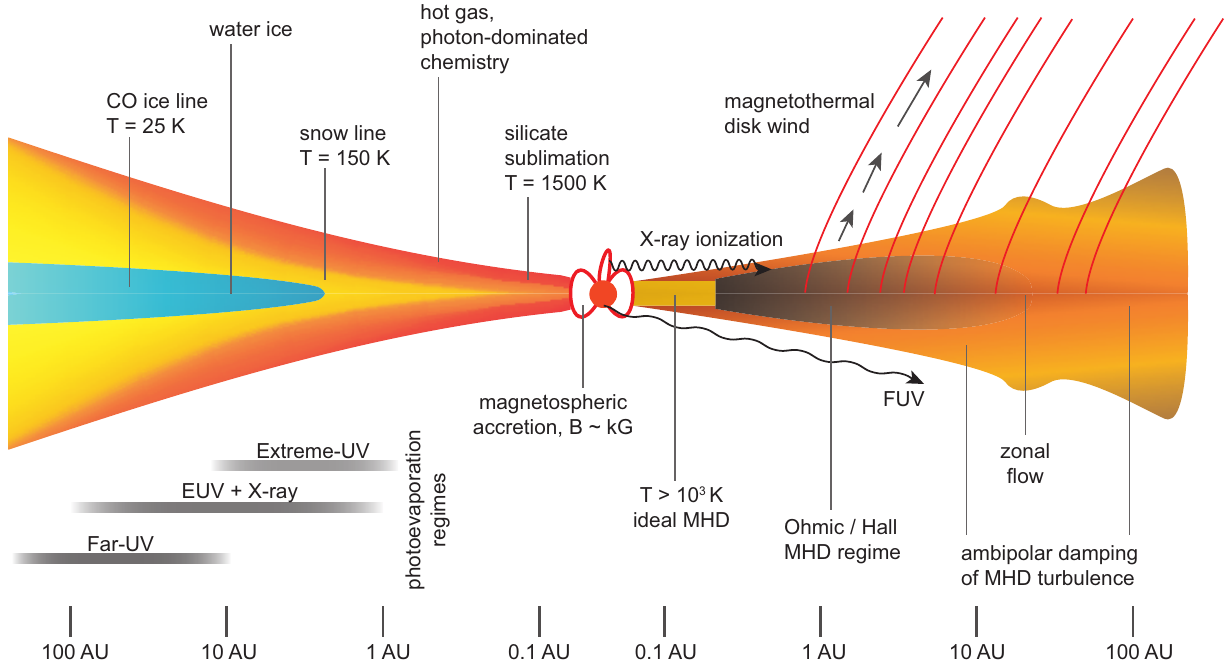}
\caption{Illustration of the thermal and ionization structure of protoplanetary disks, and the predicted 
consequences for magnetohydrodynamic (MHD) transport of angular momentum.}
\label{figure_thermal_transport}
\end{figure}

Figure~\ref{figure_thermal_transport} shows a cartoon version of disk structure. In the ``vertical" 
direction (perpendicular to the disk plane) the profile of the gas density $\rho$ is determined by 
a hydrostatic balance between the gradient of pressure $P$ and the vertical component of stellar 
gravity $g_z$,
\begin{equation}
 \frac{{\rm d}P}{{\rm d z}} = - \rho g_z.
\end{equation} 
Protoplanetary disks are observed to be thin, in that their vertical thickness is a modest 
fraction of the distance to the star, and hence we can approximate $g_z \simeq \Omega^2 z$, 
where $\Omega = \sqrt{GM_* / r^3}$ is the Keplerian angular velocity. For an isothermal gas the 
pressure is given in terms of the sound speed $c_s$ via $P=\rho c_s^2$, and the above equation 
is easily solved. An isothermal thin disk has a gaussian density profile, $\rho(z) \propto \exp (-z^2 / 2 h^2)$, with a scale height $h=c_s / \Omega$. 
In the radial direction force balance,
\begin{equation}
 \frac{v_\phi^2}{r} = \frac{GM_*}{r^2} + \frac{1}{\rho} \frac{{\rm d}P}{{\rm d}r},
\label{eq_force_balance} 
\end{equation} 
implies an orbital velocity $v_\phi = v_K [ 1 - \mathcal{O} (h/r)^2]$ that is close to Keplerian, though pressure 
support leads to a slight deviation, typically by tens of meters per second in the sense of 
sub-Keplerian rotation. This sub-Keplerian rotation has important consequences for particle dynamics.

Disks are heated by stellar irradiation and by dissipation of potential energy as gas accretes. 
Irradiation leads to a temperature profile roughly given by $T(r) \propto r^{-1/2}$ \citep{kenyon87}, 
and a disk that flares. The vertical structure in irradiation-dominated regions has 
an isothermal interior in which $T_{\rm dust} = T_{\rm gas}$, a warm layer 
of surface dust directly exposed to starlight \citep{chiang97}, and a hot gas atmosphere with photon-dominated 
chemistry. At small radii (typically at AU scales) accretion heating becomes more important, 
producing higher temperatures and replacing the isothermal interior with one in which $T(z)$ 
decreases with height. In the simple limit of radiative transfer of energy and heating in a narrow 
mid-plane slice the ratio of central to effective temperatures depends on the optical depth via 
$T_{\rm c} / T_{\rm eff} \simeq \tau^{1/4}$ \citep[e.g.][]{armitage10}, and the mid-plane is 
substantially hotter than a non-accreting disk. Accretion heating is needed to reproduce 
the location of the water snow line (at $T \simeq 150 \ {\rm K}$) in the Solar System, which 
is inferred from meteoritic evidence to have fallen at $r \approx 2.7 \ {\rm AU}$. The radius 
of the snow line changes over time as the importance of accretion heating wanes 
\citep[moving inside 1~AU at low accretion rates;][]{garaud07}, so its observed location 
in the Solar System suggests that the bodies in the asteroid belt formed relatively 
early.  Critically, the positioning of the snow line in the asteroid belt implies 
that Earth did not acquire its water in situ \citep{morbidelli00}. 

The radial distribution and evolution of the gas defy simple predictions. Dust continuum 
observations in Ophiuchus \citep[at $r \geq 20 \ {\rm AU}$ scales;][]{andrews09} and $^{13}$C$^{18}$O line 
emission from TW~Hya \citep[at 5-20~AU;][]{zhang17} suggest a surface density profile 
$\Sigma \propto r^{-0.9}$, but this cannot be predicted from first principles. 
Disk initial conditions are set by the angular momentum distribution of the collapsing cloud, 
while evolution can occur due to turbulent torques (either fluid or magnetohydrodynamic), 
large-scale laminar torques, and either thermal or 
magnetohydrodynamic (MHD) winds. In the turbulent case the disk evolves as if it has a 
large kinematic viscosity $\nu$, and it is conventional to express the efficiency of the transport 
by a dimensionless Shakura-Sunyaev $\alpha$ parameter, defined via,
\begin{equation}
 \nu = \alpha c_s h.
\end{equation} 
Order of magnitude estimates suggest that values of $\alpha = 10^{-3}-10^{-2}$ would 
suffice to drive significant disk evolution on Myr time-scales. 

Self-gravity may be the dominant 
angular momentum transport agent at early times, when the disk is massive and the Toomre $Q$ 
parameter $Q=c_s \Omega / \pi G \Sigma$ that describes the linear stability of a disk \citep{toomre64} 
is low ($Q \ale 1$). Self-gravitating disks can fragment---either when cooling of an isolated disk 
is too rapid \citep{gammie01,rice05} or when an embedded disk is over-fed with mass \citep{kratter10}---but 
this process is now considered unlikely to form a significant population of planets; the unstable radii and resultant masses are 
both predicted to be too large \citep{kratter16}. 

At later times, as the disk mass drops, MHD transport due to the 
magnetorotational instability \citep{balbus98}, the Hall shear instability \citep{kunz08}, and MHD disk 
winds \citep{blandford82,pudritz86}, is likely to dominate. Except in the innermost disk, thermally ionized at $T \age 10^3 \ {\rm K}$, 
the available sources of non-thermal ionization (X-rays, UV photons, and possibly cosmic rays if they 
are not screened) are weak enough that non-ideal MHD processes are important. There are three 
non-ideal effects \citep{wardle99}:
\begin{itemize}
\item
{\em Ohmic diffusion}, in the regime where frequent collisions couple the charged species (ions, electrons, and 
possibly charged grains) and the magnetic field to the neutrals, but there is finite conductivity.
\item
{\em Ambipolar diffusion}, where the charged species are tied to the magnetic field, but less frequent 
collisions allow the neutrals to drift relative to the field.
\item
{\em The Hall effect}, when electrons are well-coupled to the field but ions are decoupled due to 
collisions with neutrals.
\end{itemize}
The relative importance of these effects depends on location within the disk \citep[for a review, see][]{armitage11}. 
Ambipolar diffusion provides strong damping under 
the low density conditions of the outer disk ($r \age 30 \ {\rm AU}$), where a weak net vertical 
magnetic field is needed to stimulate any significant transport \citep{simon13}. At the higher densities 
on AU-scales the Hall and Ohmic terms are controlling. The action of the Hall term depends upon the 
{\em sign} of the net field with respect to the disk's rotation, and depending upon the polarity either a 
quiescent solution resembling the \citet{gammie96} dead zone, or an accreting solution driven by 
laminar MHD torques, is possible \citep{lesur14,bai14,simon15,bethune17,bai17}. The same net fields that play a 
major role in setting the level of ambipolar and Hall-dominated transport also support MHD winds, 
carrying away both mass and angular momentum \citep{bai13,gressel15}. Photoevaporative winds 
allow surface gas, heated by X-ray or UV photons, to escape at radii where $c_s \age v_K$. Photoevaporation 
alone can disperse disks on reasonable time scales \citep{alexander14}, 
though if net magnetic flux remains at late times hybrid winds driven by thermal and magnetic forces 
are expected \citep{bai16}.

Elements of this rather complex picture find observational support, though not yet highly constraining tests. 
\citet{tobin16} observe the spiral structure characteristic of gravitational instability, and fragmentation 
(into stars), in the L1448~IRS3B system. \citet{flaherty15,flaherty17}, analyzing molecular line 
profiles from the HD~163296 disk, show that turbulence is weak on scales where ambipolar 
damping would be a strong effect. Finally, a variety of studies find evidence for disk winds 
\citep{simon16}, though discrimination between thermal and MHD wind solutions is difficult. Open theoretical 
questions include the role of hydrodynamic instabilities---the most important of which may be the Vertical 
Shear Instability \citep{nelson13}---which would provide a baseline level of turbulence in magnetically dead regions. The strength and 
evolution of net disk magnetic fields 
arising from star formation is another difficult open problem.

\section{Aerodynamically controlled collisional growth}
The growth of particles from $\mu {\rm m}$ sizes up to scales of at least mm occurs almost 
everywhere within the disk via adhesive 2-body collisions. \citep[A possible exception is near ice 
lines, where vapor condensation can be competitive;][]{ros13}. The rates and outcomes of growth 
in this regime are set by aerodynamic and material physics considerations that are reasonably 
well understood.

Key to understanding the aerodynamic evolution of solid particles in disks is the realization that, 
almost always, the particles are smaller than the mean free path of gas molecules. This means 
that drag occurs in the Epstein regime, with a drag force that is linear in the relative velocity $\Delta {\bf v}$ 
between particle and gas,
\begin{equation}
 {\bf F}_{\rm drag} = - \frac{4 \pi}{3} \rho s^2 v_{\rm th} \Delta {\bf v}.
\end{equation} 
Here $\rho$ is the gas density, $v_{\rm th}$ is the thermal speed of molecules, and we have assumed 
that particles are spheres of radius $s$, mass $m$, and material density $\rho_m$ (more realistically, they would be irregular aggregates of 
small monomers). Because of the linearity, we can define a {\em stopping time} 
$t_s \equiv m \Delta v / |  {\bf F}_{\rm drag} |$ that expresses the strength of the aerodynamic 
coupling and which depends only on basic particle and gas properties, $t_s = (\rho_m / \rho)(s / v_{\rm th})$. 
Often, the physical quantity that matters most is a dimensionless version of the stopping time, 
\begin{equation}
\tau_s \equiv t_s \Omega, 
\end{equation}
obtained by multiplying through by an angular frequency (which might be the 
Keplerian frequency, or the turnover frequency of a fluid eddy). $\tau_s$ is also known as the Stokes number.

Aerodynamic forces have both local and global effects on particle evolution. Locally, the aerodynamic 
coupling of particles to turbulence (on small scales where we expect a universal Kolmogorov description 
to be valid) largely determines collision velocities, which peak for $\tau_s \sim 1$ at $\sim \sqrt{\alpha} c_s$ 
\citep{ormel07,johansen14}. Globally, aerodynamic effects lead to 
vertical settling and radial drift. Vertical settling is opposed by any intrinsic turbulence in the gas, leading 
to an equilibrium thickness of the particle disk given approximately  
by $h_d / h \simeq \sqrt{\alpha / \tau_s}$ \citep{dubrulle95}. In the absence of turbulence, 
particles in principle settle until either vertical shear ignites the Kelvin-Helmholtz instability \citep{cuzzi93}, 
or until conditions become favorable for the streaming instability \citep{youdin05}. Simultaneously, particles 
drift radially because of the slightly non-Keplerian gas rotation profile (equation~\ref{eq_force_balance}). 
For $\tau_s \ll 1$ one can think of this drift as being due to the unbalanced radial force felt by  
tightly coupled particles forced to orbit at a non-Keplerian velocity, whereas for $\tau_s \gg 1$ one 
thinks instead of a boulder orbiting at Keplerian speed and experiencing a headwind or tailwind from the 
non-Keplerian gas. In the general case, if the gas has orbital velocity $v_\phi = (1-\eta)^{1/2} v_K$ and 
radial velocity $v_{\rm r,gas}$, the particle drift speed is \citep{takeuchi02},
\begin{equation}
 v_r = \frac{\tau_s^{-1} v_{\rm r, gas} - \eta v_K}{\tau_s + \tau_s^{-1}}.
\end{equation} 
For typical disk parameters drift can be rapid, peaking at $\tau_s = 1$ where the drift 
time scale $r / | v_r|$ is only $\sim 10^3$ orbits. The direction of drift is inward if 
${{\rm d} P} / {{\rm d}r} < 0$, because in this (usual) case the radial gas pressure 
gradient partially supports the gas against gravity leading to sub-Keplerian rotation. 
Inverting this argument, however, one finds that ${{\rm d} P} / {{\rm d}r} > 0$ would 
lead to outward drift, and hence it is possible to slow or avert inward loss of solids 
in disks that have local pressure maxima. Absent such effects 
particles with $\tau_s \age 10^{-2}$ (roughly of mm-size and larger) 
are expected to drift inward and develop a time-dependent surface density profile that differs from 
that of the gas \citep{youdin04}. Andrews \& Birnstiel's chapter in this volume discusses these 
effects in detail.

The material properties of aggregates mean that some combinations of particle masses $(m_1,m_2)$ 
and collision velocities $\Delta v$ lead to bouncing or fragmentation rather than growth. If---given 
some physically plausible distribution of particle masses and collision speeds---{\em no net growth} 
occurs beyond some mass we speak of a barrier to coagulation. The existence of barriers 
is material-dependent because, at a microscopic level, the forces required to separate or rearrange 
aggregates differ for, e.g. ices and silicates \citep{dominik97}. For aggregates of $\mu {\rm m}$-sized 
silicates experiments suggest that a fragmentation barrier sets in for $\Delta v \age 1 \ {\rm m \ s^{-1}}$, while 
bouncing may set in at lower velocities \citep{guttler10}. Water ice aggregates may be able to grow 
in substantially more energetic collisions, up to at least $\Delta v \sim 10 \ {\rm m \ s^{-1}}$ \citep{gundlach15,wada09}.

The known barriers do not preclude growth up to at least mm-sizes, and models predict the 
rapid establishment of a coagulation-fragmentation equilibrium for ${\mu} {\rm m} \ale s \ale {\rm mm}$ 
in which most of the mass is in large particles \citep{birnstiel11}. At $s \sim {\rm mm}$ radial 
drift is already important, especially in the outer regions of the disk, and hence 
the gas-to-dust ratio will change as a function of radius and time. Beyond the snow line, particles plausibly 
grow until their growth time scale matches the local radial drift time \citep[``drift-limited growth";][]{birnstiel12}.  
This is due both to the intrinsic propensity of icy particles to grow to larger sizes, and to the fact that radial 
drift becomes significant at smaller physical sizes in the low density gas further out. 
In the inner disk the greater fragility of silicates means that growth may instead be frustrated by 
bouncing or fragmentation at mm-cm scales.

Multi-wavelength observations of resolved disks support part of the above picture \citep{tazzari16}, 
suggesting a radius-dependent maximum particle size in the cm (close to the star) to mm range (further out). 
There is more tension between observations and models of radial drift, with models of drift in smooth disks 
predicting faster depletion of mm-sized grains than is observed \citep{pinilla12}. Indeed, although 
the radial extent of resolved dust disks often appears markedly smaller 
than that of gas disks \citep[e.g. in TW~Hya;][]{andrews12}, detailed modeling of dust evolution and 
disk thermochemistry is needed to reliably infer the true radial variation of the dust to gas ratio \citep{facchini17}. 
The observed outer radius of a gas disk in $^{12}$CO, for example, varies substantially with $\alpha$ (which is 
not normally known), and there is a strong coupling between turbulence levels, particle sizes, and gas 
temperature. From analysis of meteorites, 
the fact that chondritic meteorites are largely made up of 
{\em chondrules}---0.1-1~mm-sized spheres of rock that were once molten---is pertinent and could be 
taken to imply a preferred size-scale for Solar Nebula solids in the asteroid belt. This interpretation is, however, 
model-dependent, and chondrule formation may involve processes 
(e.g. planetesimal collisions) unrelated to primary particle growth \citep[for a review see, 
e.g.][]{connolly16}.

The possibility of feedback loops that couple disk chemistry to disk dynamics requires further 
investigation. It is easy to sketch out a number of possible feedback mechanisms. The ionization state, 
for example, can depend sensitively upon the abundance of small dust grains (which soak up free 
charges), and may in turn determine the level of turbulence driven by MHD processes. A disk rich 
in small dust grains could then promote a low level of turbulence, rapid settling, and efficient 
coagulation. The depletion of dust might then trigger stronger levels of turbulence, and enhanced 
fragmentation, potentially 
leading to a limit cycle. Ideas in this class are physically possible, but it remains to be seen whether 
the details of disk chemistry and physics work in such a way as to realize them in disks. 

\section{Planetesimal formation}
Bridging from the aerodynamically dominated regime of mm-sized particles to gravitationally dominated 
km-scale {\em planetesimals} poses dual challenges. Growth must be fast because at intermediate scales 
where $\tau_s \sim 1$ radial drift is rapid, and must occur via a mechanism that avoids material barriers. 
No observations directly constrain the population of m-km sized bodies within primordial gas disks, 
and the observed populations of planetesimal-scale bodies (in the asteroid and Kuiper belts, and in 
debris disks where larger bodies must be present to produce the observed dust) are often heavily modified 
by collisions. Assessment of planetesimal formation models thus relies on theoretical considerations 
and circumstantial evidence.

\begin{figure}[t]
\includegraphics[width=\textwidth]{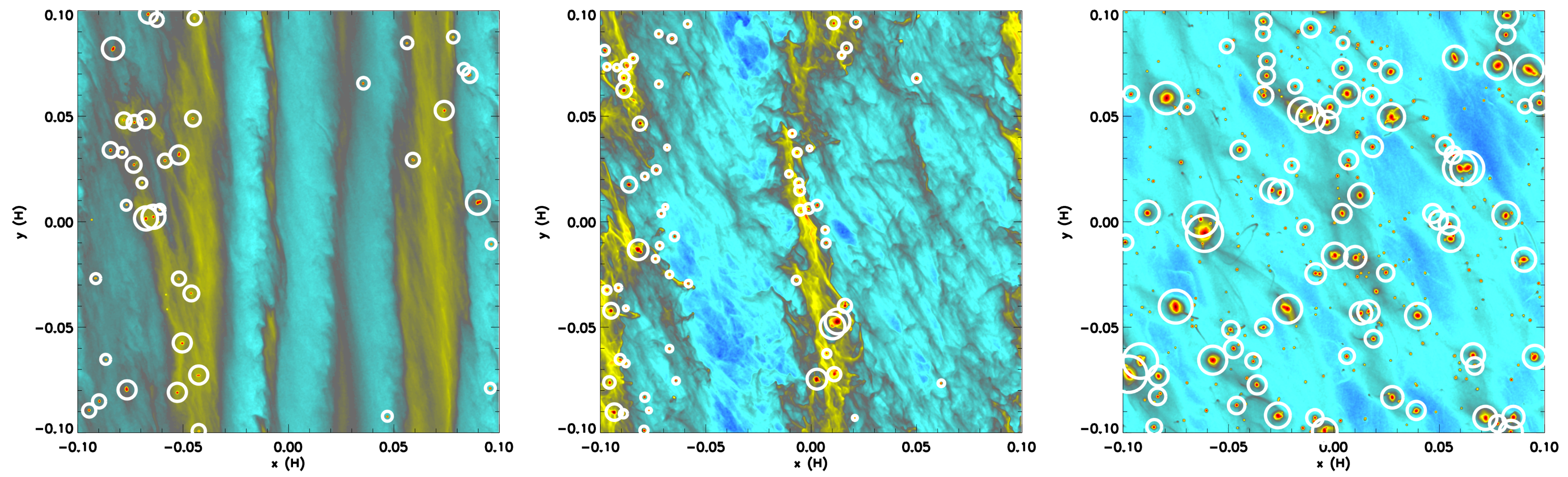}
\caption{Simulation of planetesimal formation via the gravitational collapse of streaming 
instability-induced over-densities \citep{simon17}. From left to right, the panels show the projection in the orbital plane of calculations 
run with $\tau_s = 0.006$, 0.3, and 2. The domain size was $(0.1 h)^3$.}
\label{figure_streaming}
\end{figure}

The leading hypothesis for how planetesimals form is anchored by one of the most surprising and 
consequential theoretical discoveries of recent years, the 
{\em streaming instability} \citep{youdin05}. The streaming instability is a linear instability of aerodynamically 
coupled mixtures of particles and gas that leads to small-scale 
clustering of the solids (generally on scales $\ll h$). Although the physical interpretation is 
maddeningly subtle, the instability is robust across a broad range of stopping times and 
dust-to-gas ratios, with growth time scales that are substantially longer than 
dynamical but still faster than radial drift.

The streaming instability could play a role in the collisional growth of planetesimals (if there 
are no insurmountable material barriers), but the most direct channel relies on clustering 
that is strong enough to locally exceed the Roche density, $\rho \sim M_* / r^3$. 
Simulations suggest that this strength of clustering is possible but not necessarily trivial 
to attain, requiring a minimum dust-to-gas ratio that is a function of $\tau_s$ \citep{carrera15,yang16} 
but always greater than the fiducial disk value of 0.01 \citep{johansen09}. 
Exceeding the Roche density allows clumps of 
relatively small (mm-cm) particles to gravitationally collapse, bypassing entirely the problematic scales where 
radial drift is rapid and material barriers lurk. The resulting initial mass function of 
planetesimals can be fit by a truncated power-law, ${\rm d}N / {\rm d}M \propto M^{-1.6}$ 
\citep{johansen12,simon16b,schafer17}, whose slope appears to be independent of the 
size of the particles participating in the instability \citep[Figure \ref{figure_streaming};][]{simon17}. 
This is a top-heavy mass function with most of the mass in the largest bodies. Their size in the 
inner disk, for reasonable estimates of disk properties, could be comparable to large asteroids.

Solar System constraints on streaming-initiated planetesimal formation are inconclusive. 
No observed small body population has the shallow slope that results from a single 
burst of planetesimal formation via streaming, though \citet{morbidelli09} 
argue that the size distribution of the asteroid belt is consistent with large primordial 
planetesimals and \citet{nesvorny10} suggest that gravitational collapse could 
explain the high binary fraction among classical Kuiper Belt Objects. A potentially 
important consequence of large planetesimals arises because they suffer less aerodynamic damping than small ones, leading 
to less efficient gravitational focusing and slower growth of giant planet cores in the 
classical (planetesimal dominated) formation scenario \citep{pollack96}.

The role of large scale disk structure in growth through to planetesimals is not clear. 
Several flavors of structure are observed in disks, including axisymmetric rings 
\citep{HLTau15,andrews16,isella16}, spiral arms \citep{perez16}, and horseshoe-shaped 
dust structures \citep{vdm13}. These structures could be related to zonal flows, self-gravitating 
spiral arms, and vortices, which may develop spontaneously in gas disks and which 
trap particles \citep{johansen09b,bethune17,rice06,barge95}. If this interpretation is 
right, large scale structure could be a critical pre-requisite to attaining 
conditions conducive to planetesimal formation. Alternatively, however, 
the same observed structures might be {\em caused} by planets. The planet hypothesis 
is most compelling in the case of horseshoe-shaped structures \citep[e.g.][]{zhu14}, but 
both possibilities are likely realized in nature.

\section{Terrestrial and giant planet formation}
Once planetesimals have formed, the outcome of collisions depends upon 
the energy or momentum of impacts relative to their strength---set by material 
properties for $s \ale {\rm km}$ and by gravity thereafter. Scaling laws derived 
from simulations \citep{leinhardt12} can be used as input for N-body simulations. 
Collisions lead to accretion if the planetesimals are dynamically cold, while 
high velocity impacts in dynamically excited populations (the current asteroid 
belt, debris disks, etc) lead to disruption and a collisional cascade 
that grinds bodies down to small particles \citep[in the simplest case, with a size 
distribution $n(s) \propto s^{-7/2}$;][]{dohnanyi69}.

The classical model for forming protoplanets and giant planet cores assumes that 
growth occurs within an initially cold disk of planetesimals. Consider a body of 
mass $M$, radius $R$, and escape speed $v_{\rm esc}$, embedded within a 
disk of planetesimals that has surface density $\Sigma_p$ and velocity dispersion 
$\sigma$. The eccentricity and inclination of the planetesimals are given by 
$e \sim i \sim \sigma / v_K$, so the disk thickness is $\sim \sigma / \Omega$. 
In the dispersion dominated regime (i.e. ignoring 3-body tidal effects) elementary 
collision rate arguments yield a growth rate \citep{lissauer93,armitage10},
\begin{equation}
 \frac{{\rm d}M}{{\rm d}t} = \frac{\sqrt{3}}{2} \Sigma_p \Omega \pi R^2 
 \left( 1 + \frac{v_{\rm esc}^2}{\sigma^2} \right).
\label{eq_pib} 
\end{equation}
The term in parenthesis describes the effect of {\em gravitational focusing}. It can 
vary by orders of magnitude, and hence growth in the classical picture is essentially 
controlled by the evolution of $\sigma$. Two regimes can be identified,
\begin{itemize}
\item
A small growing body does not affect the velocity dispersion 
(typically, $\sigma$ is set by a balance between excitation by 
planetesimal-planetesimal scattering encounters and aerodynamic damping). 
Let us assume that $\sigma = {\rm const} \ll v_{\rm esc}$. Then for two bodies in the 
same region of the disk, with masses $M_1 > M_2$, equation~(\ref{eq_pib}) gives 
${\rm d} (M_1/M_2) / {\rm d}t \propto (M_1/M_2)(M_1 ^{1/3} - M_2^{1/3} ) > 0$. Any initially 
small mass differences are amplified. This is the {\em runaway growth} phase \citep{greenberg78}.
\item
Eventually the fastest growing bodies start to excite the velocity dispersion of 
planetesimals in their immediate vicinity, slowing their own growth and allowing 
their radial neighbors to catch up. A number of {\em planetary embryos} then 
grow at comparable rates in a phase of {\em oligarchic growth} \citep{kokubo98}.
\end{itemize}
If there is no migration the outcome of these phases is a system of 
protoplanets on near-circular orbits. The dynamical stability of the system 
against planet-planet perturbations that drive orbit crossing and collisions is 
determined, approximately, by the planetary separation measured in units 
of the Hill radius. The Hill radius is defined for a planet of mass $M$, orbiting 
at distance $a$, as,
\begin{equation}
r_H = \left( \frac{M}{3M_*} \right)^{1/3} a.
\end{equation}
Physically, it specifies the distance out to which the planet's gravity dominates 
over the tidal gravitational field of the star. In the context of Solar System 
terrestrial planet formation we expect the initial growth phases to lead to 
a system of protoplanets separated by 5-10 Hill radii, 
with a similar amount of mass surviving in planetesimals. Simulations based on these 
initial conditions go on to form plausible analogs of the Solar System's terrestrial planets on 
$\sim 100$~Myr time scales \citep{chambers98,raymond09}.
 
Extension of this model to giant planet formation is conceptually 
straightforward \citep{pollack96}. Beyond the snow line growth is faster and planetary 
cores can readily reach masses in excess of $M_\oplus$. If we continue to ignore 
migration, one possible limit to growth is the finite supply of nearby planetesimals. 
A growing core can perturb planetesimals onto orbit-crossing trajectories within 
an annulus $\Delta a$ whose width scales with the Hill radius, $\Delta a = C r_H$, 
with $C$ a constant. The mass in planetesimals within this {\em feeding zone} 
is $2 \pi a \times 2 \Delta a \times \Sigma_p$, where $\Sigma_p$ is the surface 
density in planetesimals. This reservoir of planetesimals increases with the 
planet mass, but only weakly due to the $M^{1/3}$ dependence of $\Delta a$ on 
mass. Growth ceases when the planet reaches the {\em isolation mass}, when 
the mass of the protoplanet equals the mass of planetesimals in the feeding zone. 
A simple calculation shows that $M_{\rm iso} \propto \Sigma_p^{3/2} a^3$, so this 
consideration favors growth to larger masses at greater orbital radii. At radii 
beyond about 10~AU, however, scattering dominates over accretion, and it 
becomes increasingly hard to build large cores through planetesimal accretion.

Once the mass of a planetary core reaches a few $M_\oplus$ it can bind a hydrostatic 
gas envelope, forming a planet that resembles an ice giant. Above some critical core mass---probably in the $M_{\rm core} \approx 
5-20 \ M_\oplus$ range---hydrostatic envelope solutions cease to exist \citep{mizuno80}. 
Disk gas can thereafter be accreted rapidly, forming a gas giant planet. How the later 
stages of core accretion work depends in detail on how the envelope cools \citep[by convection 
and radiative diffusion;][]{rafikov06,piso15}, and is uncertain because the appropriate opacity is 
poorly known. A floor value to the opacity is provided by the value calculated for dust-free gas, 
and adopting this value minimizes the time scale for forming a gas giant. A much larger opacity 
is possible if the envelope contains grains with the size distribution inferred for the interstellar 
medium, though coagulation (either in the disk, or in the envelope itself) can reduce 
the opacity of even dusty gas by a large factor \citep{podolak03}. Classical (i.e. planetesimal 
dominated) models show that giant planet formation is possible on Myr time scales at 
3-10~AU \citep{movshovitz10}. 

Classical models involve two key assumptions whose validity has been challenged by recent work. 
The first is that that planetesimal formation consumes most or all 
of the disk's solid inventory. This is false; observations show that significant masses of small solids, 
observable at mm wavelengths, are present whenever there is evidence for a gas disk. Due to radial 
drift these particles approach growing planets with some velocity $\Delta v$, and can be captured. 
The resulting growth rate can be large, with an optimal limit in which a substantial fraction of particles 
entering the Bondi radius $r_B \equiv GM / \Delta v^2$ are accreted \citep{ormel10,lambrechts12}. 
{\em Pebble accretion} can be substantially faster than planetesimal-driven growth, depending upon 
the mass and size of surviving pebbles and on the dynamics of planets and planetesimals \citep{levison15}. 
The second assumption is that the core binds a static envelope that extends out to either the gaseous 
Bondi radius $r_{B,gas} \equiv GM / c_s^2$ or to the Hill sphere. Simulations, however, show 
that three dimensional flows continually cycle gas into and out of the region that is assumed 
to be bound in one dimensional models \citep{dangelo13,ormel15,lambrechts17}. These flows 
affect the thermodynamics of the outer envelope, and will alter the growth tracks of ice giants 
and mini-Neptunes embedded within gaseous disks.

Migration is the wildcard in most planet formation models. Gas disk migration occurs because 
of gravitational torques between planets and the disk \citep{goldreich79}, exerted at corotation and Lindblad resonances. 
In the {\em Type I} regime, relevant to planet masses $M \ale 10 \ M_\oplus$, the disk surface 
density is only weakly perturbed and the torque scales as $T \propto M^2$ (hence, the migration 
time scale $\propto M^{-1}$). The Lindblad torque is proportional to surface density, weakly 
dependent on gradients of density or temperature, and in isolation would invariably  lead to 
inward planet migration \citep{ward97}. The corotation torque, by contrast, is a complex 
function of the disk's structure and thermodynamics \citep{paardekooper11}. 
It can more than offset the Lindblad torque, leading to outward migration. 
Crucially, the two components of the torque have different dependencies on the 
disk structure, and while the sum may coincidentally cancel at one or a few locations within 
the disk \citep{hasegawa11,bitsch14} it is not generally zero. Type I migration will therefore be important whenever planets 
grow to masses of at least $0.1 - 1 \ M_\oplus$ while the gas is still present. The Solar System's 
terrestrial planets grow slowly enough to avoid migration (this would not be true for 
similar mass planets around lower mass stars), but giant planet cores 
and {\em Kepler} systems of super-Earths and mini-Neptunes will inevitably be affected. This line 
of reasoning favors models in which planet cores form at migration null points \citep{hasegawa11,hellary12,cossou14}, 
and those in which a substantial fraction of {\em Kepler} multi-planet systems were once in 
resonant configurations that later break \citep{goldreich14}.

Planets with masses of $M \age 3 (h/r)^3 M_*$ and above can start to open a gap in the disk. 
Once a gap forms migration occurs in the {\em Type II} regime, at a rate that depends upon 
the disk's evolution and on how rapidly the planet accretes \citep{durmann17}. The existence 
of resonant pairs of massive extrasolar planets provides strong evidence for the importance 
of this flavor of migration \citep{lee02}. The argument is simple: a pair of massive planets 
is unlikely to either form or be scattered into a resonant configuration \citep{raymond08}, because 
such configurations are a small subset of stable orbital elements. Convergent Type~II migration, 
however, can readily form resonant systems because initially well-separated planets have a 
non-zero probability of becoming locked when they encounter 
mean motion resonances \citep{goldreich65}.

\section{Long term evolution of planetary systems}
The physical processes outlined above are not fully understood, but 
even if they were there would still be considerable freedom in chaining them together to make a 
complete planet formation model. (Therein lies the promise and peril of population 
synthesis models.) It is clear, however, that some generic expectations---for example that massive 
planets should have near-circular orbits---are grossly in error. This mismatch points to the importance 
of dynamical processes that reshape planetary systems between the disk 
dispersal epoch and the time at which they are observed. 

Two planet systems are unconditionally stable against close encounters if the orbital separation 
exceeds a critical number of Hill radii \citep{gladman93}, but richer planetary systems have a ``soft" 
stability boundary and are typically unstable on a time scale that is a steep function of the 
separations \citep{chambers96,obertas17}. Unstable systems of massive planets stabilize 
by physical collisions (at small radii) and ejections (further out), leaving survivors with 
eccentric orbits. If we assume that a large fraction of giant planet systems form in ultimately 
unstable configurations, simple scattering experiments show good agreement with the 
observed eccentricity distribution of massive extrasolar planets \citep{chatterjee08}. The low 
eccentricities of the Solar System's giant planets would then imply either that scattering 
never occurred, or that it was followed in the Solar System by a phase in which dynamically 
excited orbits were damped back down.

The existence of debris disks \citep{wyatt08} indicates that disks of planetesimal-scale 
bodies are common in outer planetary systems (see Wyatt's chapter in this volume). Scattering of this material 
by gas or ice giants leads to smooth orbital migration that tends to circularize orbits. 
In the Solar System, scattering of a massive primordial Kuiper Belt would have led 
to the outward migration of Neptune \citep{fernandez84}, and the capture (and excitation 
of eccentricity) of Pluto and other 
KBOs into resonance with Neptune \citep{malhotra95}. The {\em Nice Model}, discussed 
further in Morbidelli's chapter, embeds 
this evolution into a broader framework for outer Solar System evolution, 
in which the giant planets formed in a compact resonant configuration that was 
subsequently disrupted, leading to planet-planet scattering and planetesimal-driven 
migration \citep{levison11}. Similar dynamics in extrasolar planetary systems 
would reduce the eccentricities of giant planets at larger orbital radii to Solar System-like values \citep{raymond10}.

The typical star is part of a binary system. Binary companions---if misaligned to the planetary 
orbital plane---can excite large-amplitude oscillations in $e$ and $i$ via the Kozai-Lidov 
effect \citep{naoz16}. When coupled to tidal evolution, Kozai-Lidov migration 
provides an obvious (though not unique) channel for the formation of misaligned hot Jupiters \citep{wu03}.

\begin{acknowledgement}
I acknowledge support from NASA and the National Science Foundation.
\end{acknowledgement}

\bibliographystyle{spbasicHBexo}

\end{document}